\documentclass[12pt]{article}

\title{Very special relativity as particle in a gauge field and two-time physics}
\author{Juan M. Romero\thanks{jromero@correo.cua.uam.mx},  Eric S. Escobar-Aguilar\thanks{208366868@alumnos.cua.uam.mx},  
 Etelberto V\'azquez\thanks{208366169@alumnos.cua.uam.mx}
\\[0.5cm]
\it Departamento de Matem\'aticas Aplicadas y Sistemas,\\
\it Universidad Aut\'onoma Metropolitana-Cuajimalpa\\
\it M\'exico, D.F  01120, M\'exico\\[0.3cm]} 

\begin{document}

\pagestyle{plain}

\maketitle

\begin{abstract}
The action for a $(3+1)$-dimensional  particle in very special relativity  is studied. 
It is proved that massless particles only travel in effective $(2+1)$-dimensional space-time.
It is remarkable that this action can be written as an action for a relativistic particle in a background gauge field and it is shown  that this field causes  the dimensional reduction. 
A new symmetry  for this system is found. Furthermore, a  general action with restored  Lorentz symmetry is proposed for this system. It is shown that this new action contains very special relativity and two-time physics. 
\end{abstract}

\section{Introduction}
\label{s:Intro}

Lorentz symmetry has been the most important symmetry of physics for  more than a hundred years. 
However,  is very interesting to study systems with  Lorentz violation at high energy,  in fact  some results   about  these systems  are very attractive. 
For example,   Ho\v{r}ava  gravity breaks locally this symmetry,  but it  is  power counting renormalizable  \cite{horava:gnus}. Additionally,  in quantum field theory 
  extensions of the standard model  without    Lorentz invariance  have been proposed  \cite{kostelecky:gnus,anselmi1:gnus}, some of them improve the behavior of Feynman diagrams  \cite{anselmi2:gnus}. Now,  if Minkowski geometry is changed to another generalized geometry,  Lorentz symmetry is changed too. 
 Finsler geometry  was recently put forward to replace Minkowski geometry  and it  is a  background for several  systems that   break Lorentz symmetry \cite{girelli:gnus}.
 In addition,  Cohen and Glaslow  showed that using  space-time translations plus  a special Lorentz subgroup it is possible to get a simulacrum of special relativity \cite{glashow:gnus}, this new theory was named Very Special Relativity (VSR).   In VSR   many  of the consequences of Poincar\'e group  are unchanged, but  it leaves  invariant some  constant vectors, the so called "spurion fields".  Cohen and Glashow suggested that VSR  might be important at Planck scale. It is worth to mention that Finsler geometry is a good framework for VSR  \cite{finsler:gnus} and    a relation between this theory  and  non commutative spaces was found  \cite{jabbari:gnus,ghosh:gnus}.  Furthermore,   a generalized VSR was 
 presented, where  the usual line element is  changed to \cite{gibbons:gnus}  
\begin{eqnarray}
ds^{2}=\left(-dX^{\mu}dX_{\mu}\right)^{\frac{1-b}{2}}\left(-\alpha_{\mu} dX^{\mu}\right)^{b},
\label{element-line:eq}
\end{eqnarray} 
here  the vector  $\alpha_{\mu}$ is  constant.  This line element also was found in \cite{ruso:gnus}. Some works about Lorentz violation can be seen in  \cite{kiritsis:gnus,bola:gnus,bola1:gnus,bola2:gnus,bola3:gnus,bola4:gnus,bola5:gnus,bola6:gnus,bola7:gnus,bola8:gnus,bola9:gnus,bola10:gnus,bola11:gnus,bola12:gnus,bola13:gnus,bola14:gnus,bola15:gnus}  and  works about VSR can be found  in 
\cite{grupo:gnus,grupo1:gnus,grupo2:gnus,grupo3:gnus,grupo4:gnus,grupo6:gnus,grupo7:gnus,grupo8:gnus}.\\

In this work  it will be shown  that massless particles in VSR  only travel in effective (2+1)-dimensional space-time.  This is a remarkable result, in fact  there are some approaches to quantum gravity with dynamical dimensional reduction at Planck scale 
\cite{rd1:gnus,rd2:gnus,rd3:gnus}.  Furthermore we will find that the action for  a particle in VSR  can be written as an action of relativistic particle in a background gauge field,  and  this field causes  the dimensional reduction.  The symmetries of this action will be studied. Now, in order to understand breaking symmetry we can take two approaches, the first one is take a known system invariant under a symmetry group and then  add terms which are not invariant under that group.   The second one consisting in take a system without invariance under a group transformation and then impose that  invariance. The second path is interesting, for example   the action of free complex scalar field is not invariant under local $U(1)$ group, but if we impose this symmetry we get complex scalar field in electromagnetic theory. 
Then, with  the aim to restore Lorentz symmetry in VSR,  we will take the second approach. In this approach, we will  get a generalized action, which   
 contains the action of VSR  and two-time physics.  It is worth  mentioning that  two-time physics  acts like a model that unifies the dynamic of different systems 
\cite{marnelius1:gnus,marnelius2:gnus,y-tt:gnus,bars:gnus}. In particular, this theory contains the relativistic free particle and different particles in  a curved space-time \cite{bars:gnus}. Namely, at the beginning  we had   a  system with breaking  Lorentz symmetry,  and  when  this symmetry  is restored a unified model is obtained.  Then  we can conjecture that if at Planck  scale the Lorentz symmetry is broken, as long as  it is  restored a unified model is obtained too. A work about two-time physics and breaking Lorentz symmetry  can be seen in \cite{bola4:gnus}.  It is important  to mention that recently optical metamaterials 
with two-time properties have been  proposed   \cite{meta:gnus}.\\

This manuscript is organized in the following way:  In section 2 it is shown that the  action for a particle in VSR is equivalent to a relativistic particle in a background gauge field; in section 3 Lorentz symmetry is restored; in section 4 the two-time action
is obtained and finally in section 5 our summary is given.

\section{Very special relativity as a particle in a gauge field}

The  line element (\ref{element-line:eq}) implies the action 
\begin{eqnarray}
S=-m \int d\tau \left(-\dot X^{2}\right)^{\frac{1-b}{2}}\left(-\alpha \cdot \dot X\right)^{b}.
\label{vsr:eq}
\end{eqnarray}
If $\alpha\cdot \alpha=0,$ this action induces the dispersion relation
\begin{eqnarray}
p^{2}+m^{2}\left(1-b^{2}\right)\left(\frac{-\alpha\cdot p }{m(1-b)}\right)^{\frac{2b}{1+b}}=0.
\end{eqnarray}
In this work, an arbitrary vector $\alpha_{\mu}$ will be taken.\\
  
The action  (\ref{vsr:eq}) has the  following constants: $m,\alpha_{\mu}, b.$ 
The case   $b=0$ gives  the usual relativistic particle action and   $b=1$ gives 
\begin{eqnarray}
S=m \int d\tau \alpha \cdot \dot X,
\end{eqnarray}
which does not have dynamic. For those reasons, we will take  $b\not =0$ and $ b\not =1.$ Notice 
that the cases  $m=0$ or $\alpha_{\mu}=0$ can not be studied in the action (\ref{vsr:eq}). \\

The action  (\ref{vsr:eq}) is equivalent to
\begin{eqnarray}
S&=&\int d\tau\left(  \frac{-\dot X^{2}}{\lambda_{1}} + \frac{-\alpha \cdot \dot X}{\lambda_{2}^{\frac{1-b}{2b}}}
+\lambda_{1}\lambda_{2}\left(\frac{-m}{a}\right)^{\frac{2}{1-b}}\right),\label{vsr2:eq} \\
a&=&2 \left(\frac{1-b}{2b}\right)^{b}+ \left(\frac{1-b}{2b}\right)^{b-1},
\end{eqnarray}
where $\lambda_{1}$ and $\lambda_{2}$ are Lagrange multipliers. In this  alternative action  the cases $m=0$ and $\alpha_{\mu} =0$  can be studied.\\

If  $\alpha_{\mu} =0$  we get 
\begin{eqnarray}
S&=&\int d\tau\left(  \frac{-\dot X^{2}}{\lambda_{1}} +\lambda_{1}\lambda_{2}\left(\frac{-m}{a}\right)^{\frac{2}{1-b}}\right)
\label{another:eq}
\end{eqnarray}
and the Euler-Lagrange equation to $\lambda_{2}$ is 
\begin{eqnarray}
\lambda_{1}\left(\frac{-m}{a}\right)^{\frac{2}{1-b}}=0.
\end{eqnarray}
In this case,  the action (\ref{another:eq}) is equivalent to  
\begin{eqnarray}
S&=&\int d\tau \frac{-\dot X^{2}}{\lambda_{1}} ,
\end{eqnarray}
which is the usual massless particle action.\\

When $m=0$   we obtain
\begin{eqnarray}
S&=&\int d\tau\left(  \frac{-\dot X^{2}}{\lambda_{1}} + \frac{-\alpha \cdot \dot X}{\lambda_{2}^{\frac{1-b}{2b}}}\right),
\label{vsr3:eq}
\end{eqnarray}
which  is invariant under 
\begin{eqnarray}
X^{\mu}\to \Lambda X^{\mu},\qquad \lambda_{1}\to\Lambda^{2}\lambda_{1},\qquad  \lambda_{2}\to\Lambda^{\frac{2b}{1-b}}\lambda_{2}, 
\end{eqnarray}
where $\Lambda$ is a constant. In this case the system has more symmetries. The Euler-Lagrange equation for  $\lambda_{1},\lambda_{2}$ are
\begin{eqnarray}
 \dot X^{2}=0,\qquad \alpha \cdot \dot X=\dot X^{P}=0.
\end{eqnarray}
Then,   the particle travels with  velocity of light,  but it does not travel in the direction $X^{P},$  which  means that massless particles only travel  in effective $(2+1)$-dimensional space-time. Now, at high energy  regime  $p$ is bigger than $m,$  then  we can take $m\approx 0.$ Then at Planck scale it is correct to take $m=0,$ because it is a very energetic regime.
For this reason,  at this  regime  all particles are  $(2+1)$-dimensional. This is a remarkable result, in fact  there are some approaches to quantum gravity with dynamical dimensional reduction at Planck scale 
\cite{rd1:gnus,rd2:gnus,rd3:gnus}.  \\

\subsection{Gauge symmetry}

The action (\ref{vsr2:eq})  is invariant under the  transformation
\begin{eqnarray}
\alpha_{\mu}\qquad \to \qquad  \alpha_{\mu}+\lambda_{2}^{\frac{1-b}{2b}}\frac{\partial \chi}{\partial X^{\mu}},
\end{eqnarray}
where $\chi$ is an arbitrary function of space-time. This transformation   can be interpreted as a gauge symmetry.
This result allows   to  write  the action  (\ref{vsr2:eq}) as  a particle in a background gauge field. 
In fact, using 
\begin{eqnarray}
A_{\mu}=\alpha_{\mu},  
\end{eqnarray}
the action (\ref{vsr2:eq}) becomes 
\begin{eqnarray}
S=\int d\tau\left(  \frac{-\dot X^{2}}{\lambda_{1}} -q_{eff} A_{\mu} \dot X^{\mu} +\lambda_{1} \frac{m_{eff}^{2}}{4} \right)
\end{eqnarray}
here
\begin{eqnarray}
m_{eff}^{2}=4\lambda_{2}\left(\frac{-m}{a}\right)^{\frac{2}{1-b}},   \qquad q_{eff}= \lambda_{2}^{\frac{b-1}{2b}}.
\end{eqnarray}
Then, the action (\ref{vsr2:eq})  is equivalent to 
\begin{eqnarray}
S&=&- \int d\tau\left( m_{eff}\sqrt{-\dot X^{2}} +q_{eff} A_{\mu} \dot X^{\mu}  \right),
\label{gauge:eq}
\end{eqnarray}
which  looks  like  an action of a particle in a background gauge field  $A_{\mu},$ where  mass  $m_{eff}$ and charge $e_{eff}$
are function of $\tau.$  \\
 
\section{Gauge field, restoring Lorentz symmetry and a general action}

Since  $A_{\mu}=\alpha_{\mu}$ is a constant gauge field,   it does not have  dynamics. This field  is a particular case of a  general vector  
$B_{\mu}(X),$ then a generalized VRS action is given by   
\begin{eqnarray}
S&=&\int d\tau\left(  \frac{-\dot X^{2}}{\lambda_{1}} + \frac{-B \cdot \dot X}{\lambda_{2}^{\frac{1-b}{2b}}}
+\lambda_{1}\lambda_{2}\left(\frac{-m}{a}\right)^{\frac{2}{1-b}}\right).\label{vsrl:eq} 
\end{eqnarray}
If  $B_{\mu}(X)$ is a vector  under Lorentz transformations, this symmetry is restored in (\ref{vsrl:eq}).  When $B_{\mu}(X)=\alpha_{\mu},$ the Lorentz symmetry
 is broken and the action (\ref{vsrl:eq}) reduces to VSR action (\ref{vsr:eq}).\\

When  $m=0,$  we obtain
\begin{eqnarray}
S&=&\int d\tau\left(  \frac{-\dot X^{2}}{\lambda_{1}} + \frac{-B \cdot \dot X}{\lambda_{2}^{\frac{1-b}{2b}}}\right).
\label{vsr3:eq}
\end{eqnarray}
In this case the  Euler-Lagrange equation for  $\lambda_{1},\lambda_{2}$ are
\begin{eqnarray}
 \dot X^{2}=0,\qquad B \cdot \dot X=\dot X^{P}=0.
\end{eqnarray}
This implies  that   the particle travels with  velocity of light,  but it does not travel in the direction $X^{P},$  which depends on  $B_{\mu}.$
Then the dynamical dimensional reduction is ruled  by the field $B_{\mu}.$ However,  when  $b\to 0$ this field disappears  in the action (\ref{vsr3:eq}) and  
  the particle becomes $(3+1)$-dimensional.\\

 The new action (\ref{vsrl:eq}) is invariant under gauge transformation
\begin{eqnarray}
B^{\prime}_{\mu}\left(X\right)=B_{\mu}\left(X\right)+\lambda_{2}^{\frac{1-b}{2b}}\frac{\partial \chi}{\partial X^{\mu}}.
\end{eqnarray}
Using $B_{\mu},$ it is possible construct a covariant derivative in the following way 
\begin{eqnarray}
D_{\mu}=\partial_{\mu}+i\lambda_{2}^{\frac{b-1}{2b}}B_{\mu},
\end{eqnarray}
which gives the  field strength tensor   
\begin{eqnarray}
[D_{\mu}, D_{\nu}]=i\lambda_{2}^{\frac{b-1}{2b}}F_{\mu\nu}, \qquad F_{\mu\nu}=\partial_{\mu}B_{\nu}-\partial_{\nu}B_{\mu}.
\end{eqnarray}
If VSR is important at Planck scale, $F_{\mu\nu}$ might be important at that  regime too. 
However,   $F_{\mu\nu}$   vanish when $b\to 0.$ \\


Now, if we take
\begin{eqnarray}
\lambda_{2}^{\frac{1-b}{2b}}=\frac{\lambda_{1}}{2\zeta}
\end{eqnarray}
we get
\begin{eqnarray}
S&=&\int d\tau\left(  \frac{-1}{\lambda_{1}}\left( \dot X^{2}+2\zeta  \dot X\cdot B\right) 
+\frac{\lambda_{1}^{\frac{1+b}{1-b}}}{(2\zeta)^{\frac{2b}{1-b}}}\left(\frac{-m}{a}\right)^{\frac{2}{1-b}} \right)\\
&=&\int d\tau\left(  \frac{-1}{\lambda_{1}}\left( \dot X_{\mu} +\zeta  B_{\mu}\right)^{2}+ \frac{\zeta^{2} B^{2}}{\lambda_{1}} 
+\frac{\lambda_{1}^{\frac{1+b}{1-b}}}{(2\zeta)^{\frac{2b}{1-b}}}\left(\frac{-m}{a}\right)^{\frac{2}{1-b}} \right).\label{vsr4:eq}
\end{eqnarray}
This action is a particular case of 
\begin{eqnarray}
S&=&\int d\tau\left(  \frac{-1}{\lambda_{1}}\left( \dot X_{\mu} +\zeta B_{\mu}\right)^{2}- \lambda_{1}D B^{2} 
+\frac{\lambda_{1}^{\frac{1+b}{1-b}}}{(2\zeta)^{\frac{2b}{1-b}}}\left(\frac{-m}{a}\right)^{\frac{2}{1-b}} \right) \label{gen:eq},
\end{eqnarray}
in fact, if we take 
\begin{eqnarray}
D=-\frac{\zeta^{2}}{\lambda^{2}_{1}},
\end{eqnarray}
we obtain  (\ref{vsr4:eq}). Then the action (\ref{vsr4:eq}) is more general than the action for very special relativity
and its symmetries depend on $B_{\mu}.$\\

We saw  that if $m=0$ 
the action (\ref{vsr2:eq}) has more symmetries. Now, when we take $m=0$ in (\ref{vsr4:eq}),  we get  
\begin{eqnarray}
S&=&\int d\tau\left(  \frac{-1}{\lambda_{1}}\left( \dot X_{\mu} +\zeta B_{\mu}\right)^{2}- \lambda_{1}D B^{2} 
 \right),\label{two-times2:eq}
\end{eqnarray}
which is invariant under scale transformation
\begin{eqnarray}
X^{\mu}\to \Lambda X^{\mu},\quad B^{\mu}\to \Lambda B^{\mu},\quad \lambda_{1} \to \Lambda^{2} \lambda_{1},\quad D\to \Lambda^{-4} \lambda_{1},\quad \zeta\to \zeta.
\end{eqnarray}

\section{Two-time physics}

$X^{\mu}$  is a natural vector field  under Lorentz transformation. For  this field, we have 
\begin{eqnarray}
S&=&\int d\tau L= \int d\tau\left(  \frac{-1}{\lambda_{1}}\left( \dot X_{\mu} +\zeta X_{\mu}\right)^{2}- \lambda_{1}D X^{2}  \right),\label{eq:adele}
\end{eqnarray}
which is invariant under 
\begin{eqnarray}
X^{\mu}\to \Lambda X^{\mu},\quad \lambda_{1} \to \Lambda^{2} \lambda_{1},\quad D\to \Lambda^{-4} \lambda_{1},\quad \zeta\to \zeta.
\end{eqnarray}
The action (\ref{eq:adele}) is equivalent to two-time physics action, to show this statement  notice that
\begin{eqnarray}
P_{\mu}&=&\frac{\partial L}{\partial \dot X^{\mu}}= \frac{-2}{\lambda_{1}}\left( \dot X_{\mu}+\zeta X_{\mu}\right),\nonumber\\
P_{\lambda_{1}}&=&\frac{\partial L}{\partial \dot \lambda_{1}}=0,\nonumber\\
P_{\gamma}&=&\frac{\partial L}{\partial \dot \zeta}=0,\\
P_{D}&=&\frac{\partial L}{\partial \dot D}=0.
\end{eqnarray}
Then the canonical Hamiltonian is
\begin{eqnarray}
H_{c}=-\frac{\lambda_{1}}{4} P^{2}-\zeta P\cdot X+\lambda_{1} D X^{2},
\end{eqnarray}
meanwhile  the total  Hamiltonian is \cite{dirac:gnus}
\begin{eqnarray}
H_{T}=-\frac{\lambda_{1}}{4} P^{2}-\zeta P\cdot X+\lambda_{1} D X^{2}+h_{1}P_{\lambda_{1}}+ h_{2}P_{\zeta}+h_{3}P_{D},
\end{eqnarray}
where $h_{1},h_{2},h_{3}$ are Lagrange multipliers. Therefore,  using  Dirac's method \cite{dirac:gnus}
we find 
\begin{eqnarray}
\dot P_{\lambda_{1}}&=&\{P_{\lambda_{1}}, H_{T}\}= \left(\frac{P^{2}}{4}-DX^{2}\right)\approx 0,\nonumber\\
\dot P_{\zeta}&=&\{P_{\zeta}, H_{T}\}=P\cdot X \approx 0,\\
\dot P_{D}&=&\{P_{D}, H_{T}\}=\lambda_{1} X^{2}\approx 0,
\end{eqnarray}
which implies the first class constraints 
\begin{eqnarray}
\phi_{1}&=&\frac{P^{2}}{2}\approx 0,\nonumber\\
\phi_{2}&=&P \cdot X \approx 0, \\
\phi_{3} &=&\frac{X^{2}}{2}\approx 0.
\end{eqnarray}
With these constraints we obtain    the  extended Hamiltonian \cite{dirac:gnus}  
\begin{eqnarray}
H_{E}=H_{T}+ \beta_{1} \phi_{1}+ \beta_{4} \phi_{2}+\beta_{5} \phi_{3},
\end{eqnarray}
it can be written as 
\begin{eqnarray}
H_{E}=\gamma _{1} \phi_{1}+ \gamma_{2} \phi_{2}+\gamma_{3} \phi_{3},
\label{eq:2T}
\end{eqnarray}
where 
\begin{eqnarray}
\gamma_{1}= \left(\beta_{1}-\frac{\lambda_{1}}{2}\right),\qquad  \gamma_{1}= \left(-\zeta +\beta_{2}\right),\qquad  \gamma_{3}= \left(2\lambda_{1} D+\beta_{3}\right).
\end{eqnarray}
Expression (\ref{eq:2T})   is the  two-time physics Hamiltonian which is invariant under local $SL(R,2)$ and global conformal group $SO(2,d)$ \cite{bars:gnus,y-tt:gnus},  
another work about this system can be found in \cite{ruso-chileno:gnus}.\\

Notably   two-time physics  acts as  a model that unifies  dynamics of different systems \cite{bars:gnus}.  In particular, this theory contains the relativistic free particle,  the non-relativistic particle and other particles  \cite{bars:gnus}. Then, when the  Lorentz symmetry is restored in the action (\ref{gen:eq}),  a unified model is obtained. 
This result allows   conjecture that if  Lorentz symmetry is broken at Planck scale, when this symmetry is restored a unified model is obtained. A work about two-time physics and breaking Lorentz symmetry  can be seen in \cite{bola4:gnus}. It is interesting to mention that recently  have been  proposed
optical metamaterials  with two-time properties  \cite{meta:gnus}.\\

\section{Summary}

In this work the action for a particle in very special relativity  was studied. It was shown that this action can be  written as an action for a relativistic particle in a background gauge field. This gauge field causes that the massless particles become  $(2+1)$-dimensional. Now,  at Planck scale the particles are  very energetic  and we can take $m=0,$ 
for all particles. Then, at this regime all particles are  $(2+1)$-dimensional.  This is a remarkable result, in fact  there are some approaches to quantum gravity with dynamical dimensional reduction at Planck scale \cite{rd1:gnus,rd2:gnus,rd3:gnus}. In VSR the dimensional reduction is ruled by the gauge field, which  disappears  in the limit $b\to 0.$ In that limit all  particles become $(3+1)$-dimensional. Furthermore, the Lorentz symmetry was restored  and generalized action was obtained. 
It was shown that this new action contains very special relativity and two-time physics.  It is worth notice that two-time physics  acts as  a model that unifies  dynamics of different systems \cite{bars:gnus}. This  last result allows   conjecture that if  Lorentz symmetry is broken at Planck scale, when this symmetry is restored a unified model is obtained.
Then, it is possible that  at Planck scale the Lorentz symmetry is broken and  there is  dimensional reduction, but at the regime where  the Lorentz symmetry is restored a unified model is obtained.


\end{document}